\documentclass{LMCS}

\usepackage{epsf}
\usepackage{epsfig}
\usepackage{verbatim}
\usepackage{proof}
\usepackage{latexsym}
\usepackage{amssymb}
\usepackage{hyperref}
\newtheorem{theorem}{Theorem}[section]
\newtheorem{corollary}[theorem]{Corollary}
\newtheorem{lemma}[theorem]{Lemma} 
\newtheorem{definition}[theorem]{Definition}

\newcommand{\lfpi}[3]{\Pi #1{:}#2.#3}
\newcommand{\lflam}[2]{\lambda #1.#2}
\newcommand{\eqdef}{\triangleq}

\newcommand{\SN} {\mathsf{SN}}

\newcommand{\abs}[2]{\lambda #1 . #2}            %

\newcommand{\lbr}{\lbrack\!\lbrack}
\newcommand{\rbr}{\rbrack\!\rbrack}
\newcommand{\sem}[2] {\lbr #1 \rbr_{#2}}  %

\newcommand{\NUM}[1]{\overline{#1}}

\newcommand{\Bot}{\bot}     %

\newcommand{\red}[1]{[#1]}                       %

\newcommand{\0}{\mathsf{0}}
\newcommand{\s}{\mathsf{S}}

\newcommand{\nat}{\mathsf{Nat}}
\newcommand{\NNO}{\hbox{\sf N$_0$}}
\newcommand{\EXIT}{\mathsf{exit}}
\newcommand{\recnat}{\mathsf{Rec}}

\newcommand{\VEC}{\mathsf{vec}}

\newcommand{\brec}{\Phi}
\newcommand{\brecp}{\Psi}

\newcommand{\nzero}{\hbox{\sf N$_0$}}
\newcommand{\none}{\hbox{\sf N$_1$}}

\newcommand{\PLEQ}{\mathsf{less}}

\newcommand{\get}{\mathsf{get}}
\newcommand{\HEAD}{\mathsf{head}}
\newcommand{\TAIL}{\mathsf{tail}}
\newcommand{\getaux}{\mathsf{trim}}

\def\fin{\enspace\vbox{\hrule\hbox{\vrule\kern4pt
\vbox{\kern4pt\kern4pt}\kern4pt\vrule}\hrule}}

\def\DD{\hbox{\sf D}}

\def\FUN{\hbox{\sf Fun}}
\def\TOP{\top}
\def\TOT{{\sf TP}(\DD)}
\def\POW{{\sf Pow}(\DD)}

\def\SET{\hbox{\sf U}}

\def\ZERO{\mathsf{0}}
\def\SUCC{\hbox{\sf S}}
\def\INL{\hbox{\sf Inl}}
\def\INR{\hbox{\sf Inr}}
\def\PAIR{\hbox{\sf Pair}}

\newcommand{\SEM}[1]{\mbox{$\lbr #1 \rbr$}}

\newcommand{\Elem}[1]{{\sf El}~{#1}}

\def\doi{3 (4:12) 2007}
\lmcsheading%
{\doi}
{1--16}
{}
{}
{Mar.~\phantom{0}4, 2007}
{Dec.~\phantom{0}4, 2007}
{}   
\begin{document}

\title{A proof of strong normalisation using domain theory}

\author[T.~Coquand]{Thierry Coquand\rsuper a}
\address{{\lsuper a}Chalmers Tekniska H\"ogskola, Gothenburg}
\email{coquand@cs.chamlers.se}

\author[A.~Spiwack]{Arnaud Spiwack\rsuper b}
\address{{\lsuper b}LIX, Ecole Polytechnique}
\email{Arnaud.Spiwack@lix.polytechnique.fr}

\keywords{strong normalisation, $\lambda$-calculus, double-negation
shift, Scott domain, $\lambda$-model, rewriting, denotational semantics}
\subjclass{F.4.1}

\begin{abstract}
Ulrich Berger presented a powerful proof of strong normalisation using
domains, in particular it simplifies significantly Tait's proof of
strong normalisation of Spector's bar recursion.  The main
contribution of this paper is to show that, using ideas from
intersection types and Martin-L\"of's domain interpretation of type
theory one can in turn simplify further U. Berger's argument. We build
a domain model for an untyped programming language where U. Berger has
an interpretation only for typed terms or alternatively has an
interpretation for untyped terms but need an extra condition to deduce
strong normalisation.  As a main application, we show that
Martin-L\"of dependent type theory extended with a program for Spector
double negation shift is strongly normalising.
\end{abstract}

\maketitle

\section*{Introduction}
 In 1961, Spector \cite{Spector} presented an extension of G\"odel's system $T$
by a new schema of definition called bar recursion. With this new schema, he was
able to give an interpretation of Analysis, extending G\"odel's Dialectica
interpretation of Arithmetic, and completing preliminary results of Kreisel 
\cite{Kreisel}. 
 Tait proved a normalisation theorem for Spector's bar
recursion, by embedding it in a system with infinite terms \cite{Tait}. In \cite{BBC}, an
alternative form of bar recursion was introduced. This allowed to give an interpretation of Analysis
by modified realisability, instead of Dialectica interpretation. The paper \cite{BBC}
presented also a normalisation proof for this new schema, but this proof,
which used Tait's method of introducing infinite terms, was quite complex. It was
simplified significantly by U. Berger \cite{Berger,Berger1}, who used instead a modification
of Plotkin's computational adequacy theorem \cite{Plotkin}, and could prove {\em strong}
normalisation. In a way, the idea is to replace
infinite terms by elements of a domain interpretation. This domain has the property that 
a term is strongly normalisable if its semantics is $\neq\perp$ 

The main contribution of this paper is to show that, using ideas from intersection types
\cite{Akama,Bakel,Barendregt,Pottinger} and Martin-L\"of's domain interpretation of type theory
\cite{MartinLof}, one can in turn simplify further U. Berger's argument.
Contrary to \cite{Berger}, we build a domain model for an {\em untyped} programming language.
Compared to \cite{Berger1}, there is no need of an extra hypothesis to deduce
strong normalisation from the domain interpretation.
A noteworthy feature of this domain model is that it is in a natural way a {\em complete}
lattice, and in particular it has a {\em top} element 
which can be seen as the interpretation of a top-level exception in programming languages.
We think that this model can be the basis of {\em modular} proofs of strong normalisation
for various type systems. As a main application, we show that Martin-L\"of dependent
type theory extended with a program for Spector double negation shift \cite{Spector}\footnote{This is the schema
$(\forall x.\neg\neg P(x))\rightarrow \neg\neg\forall x.P(x)$. Spector \cite{Spector} remarked that it is enough
to add this schema to intuitionistic analysis in order to be able to interpret classical analysis via negative
translation.}, similar to bar recursion, has the strong normalisation property.

\section{An Untyped Programming Language}

 Our programming language is untyped $\lambda$-calculus
extended with constants, and has the following syntax. 
$$
M,N ::= x \mid \abs{x}{M} \mid M\ N \mid c \mid f
$$
There are two kinds of constants: {\em constructors}
$c,c',\dots$ and {\em defined constants} $f,g,\dots$.
We use $h,h',\ldots$ to denote a constant which may be a constructor
or defined.
Each constant has an {\em arity}, but can be partially applied.
We write ${\sf FV}(M)$ for the set of free variables of $M$.
We write $N(x=M)$ the result of substituting the free occurences of $x$ by $M$ in $N$
and may write it $N[M]$ if $x$ is clear from the context.
We consider terms up to $\alpha$-conversion.

 The computation rules of our programming language
are the usual $\beta$-reduction and $\iota$-reduction
defined by a set of rewrite rules of the form
$$
f\ p_1\ldots p_k = M
$$
where $k$ is the arity of $f$ and ${\sf FV}(M)\subseteq {\sf FV}(f\ p_1\ldots p_k)$.
In this rewrite rule,  $p_1,\dots,p_k$ are {\em constructor patterns}
\emph{i.e.} terms of the form
$$
p ::= x \mid c\ p_1\ldots p_l
$$
where $l$ is the arity of $c$. Like in \cite{Berger}, we assume our
system of constant reduction rules to be {\em left linear}, i.e. a variable
occurs at most once in the left hand side of a rule, and {\em mutually
disjoint}, i.e. the left hand sides of two disjoint rules are non-unifiable.
We write $M\rightarrow M'$ if $M$ reduces in one step to $M'$ by $\beta,\iota$-reduction
and $M=_{\beta,\iota} M'$ if $M$, $M'$ are convertible by $\beta,\iota$ conversion.
It follows from our hypothesis on our system of reduction rules that 
$\beta,\iota$-reduction is confluent \cite{KVOVR}. We write $\rightarrow (M)$ for the set of terms
$M'$ such that $M\rightarrow M'$.

We work with a given set of constants, that are listed in 
section \ref{s:application}, but our arguments are general and make use only of the fact 
that the reduction system is left linear
and mutually disjoint. We call UPL, for Untyped Programming Language, the system defined by this list
of constants and $\iota$-reduction rules. 
The goal of the next section is to define a domain model for UPL
that has the property that $M$ is strongly normalizing if $\sem{M}{}\neq\perp$.

\section{A domain for strong normalization}

\subsection{Formal Neighbourhoods}

\begin{definition}%
The {\em Formal Neighbourhoods} are given by the following grammar:
$$
U,V ::= \nabla \mid c\ U_1\ldots U_k \mid U\rightarrow V \mid U\cap V
$$
\end{definition}

On these neighbourhoods we introduce a \emph{formal inclusion} $\subseteq$ relation
defined inductively by the rules of Figure 
\ref{f:formalinclusion}. In these rules we use the formal equality relation
$U=V$ defined to be $U\subseteq V$ and $V\subseteq U$. We let $\mathcal{M}$
be the set of neighbourhoods quotiented by the formal equality.
The terminology ``formal neighbourhoods'' comes from \cite{Kreisel,Scott,MartinLof}.

\begin{figure}
$$
\begin{array}{rcl}
\nabla\cap U & = & \nabla \\
c\ U_1 \ldots U_k\cap c'\ V_1\ldots V_{l} & = & \nabla \\
c\ U_1 \ldots U_k\cap V\rightarrow W & = & \nabla \\
(U\rightarrow V_1)\cap (U\rightarrow V_2) & = & U\rightarrow (V_1\cap V_2) \\
\multicolumn{3}{c}{
\hspace{-2mm}
c\ U_1 \ldots U_k{\cap} c\ V_1 \ldots V_{k} =  c\ (U_1{\cap} V_1) \ldots (U_k{\cap} V_k)
}
\end{array}
$$
$$
\begin{array}{c@{\qquad}c}
\infer{U_1\subseteq U_3}{U_1\subseteq U_2 & U_2\subseteq U_3}
&
\infer{c\ U_1\ldots U_k \subseteq c\ V_1\ldots V_k}{U_1\subseteq V_1 & \ldots & U_k\subseteq V_k}
\\[3mm]
\infer{U\subseteq U}{}
&
\infer{U \subseteq V_1\cap V_2}{U\subseteq V_1 & U\subseteq V_2}
\\
\infer{V_1\cap V_2 \subseteq V_1}{}~~~~~\infer{V_1\cap V_2 \subseteq V_2}{}
&
\infer{U_1\rightarrow V_1 \subseteq U_2\rightarrow V_2}{U_2\subseteq U_1 & V_1\subseteq V_2}
\end{array}
$$
\caption{Formal inclusion}\label{f:formalinclusion}
\end{figure}

\begin{lemma}\label{formal} The formal inclusion and equality  are both {\em decidable} relations,
and $\mathcal{M}$ is a poset for the formal inclusion relation, and $\cap$
defines a binary meet operation on $\mathcal{M}$. We have 
$c\ U_1 \ldots U_k\neq c'\ V_1\ldots V_{l}$ if $c\neq c'$ and
$c\ U_1 \ldots U_k= c\ V_1 \ldots V_k$ if and only if $U_1=V_1, \ldots, U_k=V_k$.
An element in $\mathcal{M}$ is either $\nabla$ or of the form
$c\ U_1 \ldots U_k$ or of the form $(U_1\rightarrow V_1)\cap\ldots\cap (U_n\rightarrow V_n)$
and this defines a partition of  $\mathcal{M}$.
Furthermore the following ``continuity condition'' holds:
if $I$ is a (nonempty) finite set and $\bigcap_{i\in I} (U_i\rightarrow V_i) \subseteq U\rightarrow V$ then the set
$J = \{ i\in I \mid U\subseteq U_i \}$ is not empty and
$\bigcap_{i\in J} V_i \subseteq V$.
Note that there is no maximum element, where there usually is one. This
is linked to the fact that we are aiming to prove \emph{strong} normalisation,
not weak normalisation.
\end{lemma}

 Similar results are proved in \cite{Amadio,Akama,Barendregt,Bakel,MartinLof}.

\begin{proof}
We introduce the set of neighbourhoods in ``normal form'' by the grammar
$$
\begin{array}{lcl}
W,W' & ::= & \nabla~|~c~W_1~\dots~W_k~|~I\\
I & ::= & (W_1\rightarrow W_1')\cap\dots\cap (W_n\rightarrow W_n')
\end{array}
$$
and define directly the operation $\cap$ and the relation $\subseteq$ on this set. 
An element in normal form $W$ is of the form $\nabla$ or $c~W_1~\dots~W_k$ or is a finite formal
intersection $\cap X$ where $X$ is a nonempty finite set of elements of the form $W\rightarrow W'$.
The definition of $\cap$ and $\subseteq$ will be recursive, using the following complexity measure:
$|\nabla| = 0,~|c~W_1~\dots~W_k| = 1 + max(|W_1|,\dots,|W_k|)$ and
$|\cap _i(W_i\rightarrow W_i')| = 1 + max_i(|W_i|,|W_i'|)$.

We define 

\medskip

 $\nabla\cap W = W\cap\nabla = \nabla$

 $c~W_1~\dots~W_k\cap c~W_1'~\dots~W_k' = c~(W_1\cap W_1')~\dots~(W_k\cap W_k')$

 $c~W_1~\dots~W_k\cap c'~W_1'~\dots~W_l' = \nabla$

 $c~W_1~\dots~W_k\cap (\cap X) = (\cap X)\cap c~W_1~\dots~W_k = \nabla$
 
 $(\cap X)\cap (\cap Y) = \cap (X\cup Y)$.

\medskip

 Notice that we have $|W_1\cap W_2|\leq max(|W_1|,|W_2|).$

 We have furthermore $\nabla\subseteq W$ and $c~W_1~\dots~W_k\cap c~W_1'~\dots~W_k'$
iff $W_i\subseteq W_i'$ for all $i$ and finally $\cap X\subseteq \cap Y$ iff
for all $W\rightarrow W'$ in $Y$ there exists $W_1\rightarrow W_1',~\dots,~W_k\rightarrow W_k'$
in $X$ such that $W\subseteq W_1,~\dots,~W\subseteq W_k$ and
$W_1'\cap\dots\cap W_k'\subseteq W'$. This definition is well founded since
$|W_1'\cap\dots\cap W_k'|<|\cap X|$ and $|W'|<|\cap Y|$. One can then prove that
relation $\subseteq$ and the operation $\cap$ satisfies all the laws of  Figure \ref{f:formalinclusion}
on the set of neighbourhoods of complexity $<n$ by induction on $n$.

 Since all the laws of Figure \ref{f:formalinclusion} are valid for this structure we get in this way
a concrete representation of the poset  $\mathcal{M}$, and all the properties of this poset can be
directly checked on this representation.
\end{proof}

We associate to $\mathcal{M}$ a type system defined in Figure
\ref{f:typeinter} (when unspecified, $k$ is the arity of the related constant). 
It is a direct extension of the type systems considered
in \cite{Akama,Amadio,Bakel,Barendregt,MartinLof}. The typing
rules for the constructors and defined constants appear to be new however.
Notice that the typing of the function symbols is very close to a recursive
definition of the function itself. Also, we make use of the fact that,
as a consequence of Lemma \ref{formal},
one can define when a constructor pattern matches an element of $\mathcal{M}$.

\begin{figure}[t!]
$$
\begin{array}{c}
\infer{\Gamma \vdash_\mathcal{M} x:U}{x:U\in \Gamma}\\[3mm]
\infer{\Gamma \vdash_\mathcal{M} c: U_1\rightarrow\ldots\rightarrow U_k\rightarrow c\ U_1\ldots U_k}
      {} \\[3mm]
\infer{\Gamma \vdash_\mathcal{M} \abs{x}{M}:U\rightarrow V }
      {\Gamma, x{:}U \vdash_\mathcal{M} M:V} \\[3mm]
\infer{\Gamma \vdash_\mathcal{M} N\ M: V}
      {\Gamma \vdash_\mathcal{M} N: U\rightarrow V & \Gamma\vdash_\mathcal{M} M:U} \\[3mm]
\infer{\Gamma \vdash_\mathcal{M} M:U\cap V}{\Gamma\vdash_\mathcal{M} M:U & \Gamma\vdash_\mathcal{M} M:V}\\[3mm]
\infer{\Gamma \vdash_\mathcal{M} M:U} {\Gamma\vdash_\mathcal{M} M:V & V\subseteq U} \\[3mm]
\infer{\Gamma \vdash_\mathcal{M} f:U_1\rightarrow\ldots\rightarrow U_k \rightarrow V}
      {\begin{array}{cc}
        f\ p_1\ldots p_k = M & p_i(W_1,\ldots,W_n)= U_i \\
        \multicolumn{2}{c}{\Gamma,x_1{:}W_1,\ldots,x_n{:}W_n \vdash_\mathcal{M} M:V}
       \end{array}} \\[3mm]
\infer{\Gamma \vdash_\mathcal{M} f:U_1\rightarrow\ldots\rightarrow U_k\rightarrow\nabla}
      { \begin{array}{c}
      \textrm{for any $U_1,\dots,U_k$ such that }\\
      \textrm{no rewrite rule of $f$ matches $U_1,\dots,U_k$}
       \end{array} }
\end{array}
$$
\caption{Types with intersection in $\mathcal{M}$}\label{f:typeinter}
\end{figure}

\begin{lemma}\label{inversion1}
If $\Gamma \vdash_\mathcal{M} \abs{x}{N}: U$ then there exists a family $U_i,V_i$
such that $\Gamma,x:U_i\vdash_\mathcal{M} N:V_i$ and $\cap _i (U_i\rightarrow V_i)\subseteq U$.
\end{lemma}

\begin{proof}
Direct by induction on the derivation.
\end{proof}

\begin{lemma}\label{inversion}
If $\Gamma \vdash_\mathcal{M} \abs{x}{N}: U\rightarrow V$ then $\Gamma,x{:}U\vdash_\mathcal{M} N:V$.
\end{lemma}

\begin{proof}
We have a family $U_i,V_i$
such that $\Gamma,x:U_i\vdash_\mathcal{M} N:V_i$ and $\cap _i (U_i\rightarrow V_i)\subseteq U\rightarrow V$.
By Lemma \ref{formal} there exists $i_1,\dots,i_k$ such that $U\subseteq U_{i_1},\dots,~U\subseteq U_{i_k}$
and $V_{i_1}\cap\dots\cap V_{i_k}\subseteq V$. This together with  $\Gamma,x:U_i\vdash_\mathcal{M} N:V_i$
imply $\Gamma,x:U\vdash_\mathcal{M} N:V$.
\end{proof}

\begin{lemma}\label{NM}
If $\Gamma \vdash_\mathcal{M} N~M:V$ then there exists $U$ such that
 $\Gamma \vdash_\mathcal{M} N:U\rightarrow V$ and  $\Gamma \vdash_\mathcal{M} M:U$.
\end{lemma}

\begin{proof}
Direct by induction on the derivation.
\end{proof}

\subsection{Reducibility candidates}

\begin{definition}%
      $\mathcal{S}$ (the set of simple terms) is the set of terms that are neither an abstraction nor
      a constructor headed term, nor a partially applied destructor headed term (\emph{i.e.} 
      $f\ M_1\ \ldots\ M_n$ is simple if $n$ is greater or equal to the arity of $f$).
\end{definition}

\begin{definition}%
A {\em reducibility candidate} $X$ is a set  of terms with the following properties:
\begin{description}
\item[(CR1)] $X\subseteq\SN$
\item[(CR2)] $\rightarrow(M)\subseteq{X}$ if $M\in X$ 
\item[(CR3)] $M\in X$ if $M\in\mathcal{S}$ and $\rightarrow(M)\subseteq{X}$
\end{description}
\end{definition}

 It is clear that the reducibility candidates form a complete lattice w.r.t. the inclusion
relation. In particular, there is a {\em least} reducibility candidate $R_0$, which can be
inductively defined as the set of terms $M\in\mathcal{S}$ such that $\rightarrow(M)\subseteq R_0$. For 
instance, if $M$ is a variable $x$, then we have $M\in R_0$ since $M\in\mathcal{S}$ and
$\rightarrow(M) = \emptyset$.

We define two operations on sets of terms, which preserve the status of candidates.
If $c$ is a constructor of arity $k$ and $X_1, \ldots, X_k$ are sets of terms
then  the set $c\ X_1\ldots X_k$ is inductively defined to be the set of terms $M$
of the form $c\ M_1\ldots M_k$, with $M_1\in X_1\ldots M_k\in X_k$ or such that
$M\in\mathcal{S}$ and $\rightarrow (M)\subseteq c\ X_1\ldots X_k$.
If $X$ and $Y$ are sets of terms, $X\rightarrow Y$ is the set of terms $N$ such that
$N\ M\in Y$ if $M\in X$.

\begin{lemma}
If $X$ and $Y$ are reducibility candidates then so are $X\cap Y$ and $X\rightarrow Y$.
If $X_1,\ldots, X_k$ are reducibility candidates then so is $c\ X_1\ldots X_k$.
\end{lemma}

\begin{definition}%
The function $\red{-}$ associates a reducibility candidate to each formal neighbourhood.
\begin{itemize}
\item $\red{\nabla}\eqdef R_0$
\item $\red{c\ U_1\ldots U_k}\eqdef c\ \red{U_1}\ldots\red{U_k}$
\item $\red{U\rightarrow V}\eqdef \red{U}\rightarrow \red{V}$
\item $\red{U\cap V}\eqdef \red{U}\cap\red{V}$
\end{itemize}
\end{definition}

\begin{lemma}\label{inclusion}
If $U\subseteq V$ for the formal inclusion relation then $\red{U}\subseteq \red{V}$ as sets of terms.
\end{lemma}

This follows from the fact that all the rules of Figure \ref{f:formalinclusion} are valid 
for reducility candidates.

\begin{theorem}\label{SNinter}%
If $\vdash_\mathcal{M} M: U$ then $M\in \red{U}$. In particular $M$ is strongly normalising.
\end{theorem}

As usual, we prove that if $x_1:U_1,\ldots,x_n:U_n\vdash_\mathcal{M} M: U$ and
$M_1\in\red{U_1},\dots,M_n\in\red{U_n}$ then $M(x_1=M_1,\dots,x_n=M_n)\in\red{U}$.
This is a mild extention of the usual induction on derivations. We sketch
the extra cases:
\begin{itemize}
\item Subtyping: direct from Lemma \ref{inclusion}.
\item Constructor: direct from the definition of $\red{c\ U_1\ldots U_k}$.
\item Defined constant (case with a rewrite rule): we need a small remark:
      since $c'\ M_1\ldots M_l\not\in\mathcal{S}$ for any $l$, we have
      that $c'\ M_1\ldots M_l\in c\ X_1\ldots X_k$ implies
      $c'=c$ and $l=k$ by definition of $c\ X_1\ldots X_k$.
      Knowing this we get that if $N_i\in p_i(\red{W_1},\ldots,\red{W_n}))$,
      then $f\ N_1\ldots N_k$ can only interract with one rewrite rule
      (remember that there is no critical pair). The definition of
      $c\ X_1\ldots X_k$ also tells us that if the $N_i$ are equal
      to $p_i(M_1,\ldots,M_n)$, then $M_j\in W_j$.
      From this the result follows easily.
\item Defined constant (case with no rewrite rule): we need the same
      remark as in the previous case: 
      $c'\ M_1\ldots M_l\in c\ X_1\ldots X_k$ implies that $c'=c$ and 
      $l=k$. Additionally, $\red{\nabla}$ does not contain any 
      constructor-headed term (since $\red{\nabla}\subseteq \mathcal{S}$). 
      A consequence
      of these two remarks is that there cannot be any fully applied 
      constructor-headed term in $\red{U\rightarrow V}$, by simple induction.
      In particular there is no term matched by a pattern in 
      $\red{U\rightarrow V}$.
      Thus, since there is no rule matching the $U_1, \ldots, U_k$, we know
      that for any $N_1\in \red{U_1}, \ldots, N_k\in \red{U_k}$, 
      $f\ N_1\ldots N_k$ is not matched by any rewrite rule; it is,
      however, a simple term.
      It follows easily that $f\ N_1\ldots N_k\in\red{\nabla}$. \qed
\end{itemize}

\subsection{Filter Domain}

\begin{definition}%
An {\em I-filter}\footnote{This terminology, coming from \cite{Bakel}, stresses the fact
that the empty set is also an I-filter.} over $\mathcal{M}$ is
a subset $\alpha\subseteq \mathcal{M}$ with the following closure properties:
\begin{itemize}
\item if $U, V\in \alpha$ then $U\cap V\in\alpha$
\item if $U\in\alpha$ and $U\subseteq V$ then $V\in\alpha$
\end{itemize}
\end{definition}

It is clear that the set $\DD$ of all I-filters over $\mathcal{M}$ 
ordered by the set inclusion is a complete algebraic domain.
The finite elements of $\DD$ are exactly $\emptyset$
and the principal I-filters  $\uparrow U \eqdef \{ V \mid U\subseteq V \}$.
The element $\top = \uparrow \nabla$ is the greatest element of $\DD$
and the least element is $\perp = \emptyset$.

We can define on $\DD$ a binary application operation 
$$\alpha\ \beta \eqdef \{V \mid \exists U, U\rightarrow V\in\alpha \wedge U\in\beta \}$$
We have always  $\alpha\ \perp = \perp$ and $\top\ \beta = \top$ if $\beta\neq\perp$.
We write $\alpha_1\dots\alpha_n$ for $(\ldots (\alpha_1\ \alpha_2)\ \ldots)\ \alpha_n.$

\subsection{Denotational semantics of UPL}

 As usual, we let $\rho,\nu,\dots$ range over {\em environments}, i.e. mapping
from variables to $\DD$.

\begin{definition}%
If $M$ is a term of UPL, $\sem{M}{\rho}$ is the I-filter of neighbourhoods
$U$ such that
$x_1{:}V_1,\ldots, x_n{:}V_n \vdash_\mathcal{M} M:U$ for some $V_i\in\rho(x_i)$
with ${\sf FV}(M) = \{x_1 ,\ldots, x_n\}.$
\end{definition}

 A direct consequence of this definition and of Theorem \ref{SNinter} is then

\begin{theorem}\label{SN}%
If there exists $\rho$ such that $\sem{M}{\rho}\neq\Bot$ then $M$ is strongly normalising.
\end{theorem}

 Notice also that we have $\sem{M}{\rho} = \sem{M}{\nu}$
as soon as $\rho(x)=\nu(x)$ for all $x\in {\sf FV}(M)$. Because of this we can
write $\sem{M}{}$ for $\sem{M}{\rho}$ if $M$ is closed. If $c$ is a constructor, we
write simply $c$ for $\sem{c}.$

\begin{lemma}\label{partition}
We have
$c\ \alpha_1 \ldots \alpha_k\neq c'\ \beta_1\ldots \beta_{l}$ if $c\neq c'$ and
$c\ \alpha_1 \ldots \alpha_k= c\ \beta_1 \ldots \beta_k$ if and only if $\alpha_1=\beta_1 \ldots \alpha_k=\beta_k$,
whenever $\alpha_i\neq\perp,~\beta_j\neq\perp$.
An element of $\DD$ is either $\perp$, or $\top$ or of the form $c~\alpha_1~\ldots~\alpha_k$
with $c$ of arity $k$ and $\alpha_i\neq\perp$ or is a sup of elements of the form
$\uparrow (U\rightarrow V)$. This defines a partition of $\DD$. 
\end{lemma}

\begin{proof}
Follows from Lemma \ref{formal}.
\end{proof}

As a consequence of Lemma \ref{partition},
it is possible to define when a constructor pattern matches an element of $\DD$. 
The next result expresses the fact that we have defined in this 
way a {\em strict model} of UPL. 

\begin{theorem}\label{model}
$$
\begin{array}{lcl@{\quad}l}
\sem{x}{\rho} &=& \rho(x) \\
\sem{N\ M}{\rho} &=& \sem{N}{\rho}\ \sem{M}{\rho} \\
\sem{\abs{x}{M}}{\rho}\ \alpha &=& \sem{M}{(\rho,x:=\alpha)} &\textrm{if }\alpha\neq\Bot\\
\end{array}
$$
If $f~p_1~\ldots~p_k = M$ and $\alpha_i = \sem{p_i}{\rho}$ then
$\sem{f}{}\ \alpha_1\ldots\alpha_k = \sem{M}{\rho}$.
If there is no rule for $f$ which matches $\alpha_1,\ldots,\alpha_k$
and $\alpha_1,\ldots,\alpha_k$ are $\neq\perp$ then 
$\sem{f}{}\ \alpha_1\ldots\alpha_k = \top.$
Finally, if for all $\alpha\neq\Bot$ we have
$\sem{M}{(\rho,x:=\alpha)}=\sem{N}{(\nu,y:=\alpha)}$ then
$\sem{\abs{x}{M}}{\rho} = \sem{\abs{y}{N}}{\nu}.$
\end{theorem}

\begin{proof}
The second equality follows from Lemma \ref{NM} and
the third equality follows from Lemma \ref{inversion}.
\end{proof}
 
\begin{corollary}\label{subst}
$\sem{N(x=M)}{\rho} = \sem{N}{(\rho,x=\sem{M}\rho)}$
\end{corollary}

\section{Application to Spector's Double Negation Shift}\label{s:application}

 The goal of this section is to prove strong normalisation for dependent
type theory extended with Spector's double negation shift \cite{Spector}. The version of type
theory we present is close to the one in \cite{ML}: we have a type of natural
numbers $\nat:\SET$, where $\SET$ is an universe. It is shown in \cite{ML}, using
the propositions-as-types principle, how to represent intuitionistic higher-order
arithmetic in type theory. It is then possible to formulate Spector's double
negation shift as
$$
(\Pi n:\nat.\neg\neg B~n)\rightarrow \neg\neg \Pi n:\nat.B~n
$$
where $\neg A$ is an abreviation for $A\rightarrow \nzero$ and $B:\nat\rightarrow\SET$.
Spector showed \cite{Spector} that it is enough to add this schema (Axiom F in \cite{Spector})
to intuitionistic analysis in order to be able to interpret classical analysis via a negative
translation.
We show how to extend dependent type theory with a constant of this type in such a
way that strong normalisation is preserved. It follows then from \cite{Spector}
that the proof theoretic strength of type theory is much stronger
with this constant and has the strength of classical analysis.

\subsection{General Rules of Type Theory}

 We have a constructor $\FUN$ of arity 2 and
we write $\Pi x{:}A.B$ instead of
$\FUN~A~(\lambda x.B)$, and $A\rightarrow B$ instead of $\FUN~A~(\lambda x.B)$ if $x$
is not free in $B$. %
We have a special constant $\SET$ for universe.
(We recall that we consider terms up to $\alpha$-conversion.) A {\em context} is a sequence $x_1:A_1,\dots,x_n:A_n$, where the $x_i$ are pairwise distinct.

They are three forms of judgements
$$
\Gamma\vdash A~~~~~~~\Gamma\vdash M:A~~~~~~~\Gamma\vdash
$$

 The last judgement $\Gamma\vdash$ expresses that $\Gamma$ is a well-typed
context. We may write $J~[x:A]$ for $x:A\vdash J$.

 The typing rules are in figure \ref{f:tt}%
\begin{figure}[ht] 
$$
\frac{}{\vdash}\qquad
\frac{\Gamma\vdash A}{\Gamma,x:A\vdash}
$$
\hfill\\
$$
\frac{\Gamma\vdash}{\Gamma\vdash \SET}\qquad
\frac{\Gamma\vdash A:\SET}{\Gamma\vdash A}\qquad
\frac{\Gamma,x:A\vdash B}{\Gamma\vdash \lfpi{x}{A}{B}}
$$
\hfill\\
$$
\frac{(x:A)\in\Gamma~~~~\Gamma\vdash}{\Gamma\vdash x:A}\qquad
\frac{\Gamma,x:A\vdash M:B}{\Gamma\vdash \lflam{x}{M}:\lfpi{x}{A}{B}}\qquad
\frac{\Gamma\vdash N:\lfpi{x}{A}{B}~~~~\Gamma\vdash M:A}{\Gamma\vdash N~M:B[M]}
$$
\hfill\\
$$
\frac{\Gamma\vdash M:A~~~~\Gamma\vdash B~~~~A=_{\beta,\iota} B}{\Gamma\vdash M:B}
$$
\hfill\\
\hfill\\
 We express finally that the universe $\SET$ is closed under the product operation.
$$
\frac{\Gamma\vdash A:\SET~~~~\Gamma,x:A\vdash B:\SET}{\Gamma\vdash \lfpi{x}{A}{B}:\SET}
$$
\label{f:tt}
\caption{Typing Rules of Type Theory}
\end{figure}

 The constants are the ones of our language UPL, described in the next subsection.

\subsection{Specific Rules}

 We describe here both the untyped language UPL (which will define the $\iota$
reduction) and the fragment of type theory that we need in order to express
a program for Spector double negation shift. 
The constant of form $(op)$ are used as infix operators.

 The constructors are $\SET,\nat,\nzero,\none,\0$ (arity 0),
$\s,\INL,\INR$ (arity 1) and $(+),(\times),\FUN,\PAIR$ (arity 2). 
To define the domain $\DD$ as in the previous sections, it is enough to know
these constructors. 

The defined constants of the language UPL are
$\VEC,\get,\getaux,T,\HEAD,\TAIL,(\leq),\PLEQ, \recnat, \neg$, 
$\EXIT, \brec, \brecp$. The arities
are clear from the given $\iota$-rules. %
From these $\iota$-rules it is then possible to interpret each of these
constants as an element of the domain $\DD$.

 At the same time we introduce these constants (constructors or
defined constants) we give their intended types.

First we have the type of natural numbers $\nat$ with two constructors:

\medskip
$
\begin{array}{||l@{~}c@{~}r}
\multicolumn{3}{l}{
\hspace{-2mm}\nat:\SET}\\
\0 & : & \nat\\
\s & : & \nat
\end{array}
$
\medskip

We also add the natural number recursor $\recnat$ so that the language
contains Heyting airthmetic:

\medskip

$
\begin{array}{||l@{\ }l@{\ }l@{\ }c@{\ }cl}
\multicolumn{6}{l}{
\hspace{-2mm}\recnat : C~\0\rightarrow (\Pi n:\nat.C\ n\rightarrow C\ (\s\ n))\rightarrow \Pi n:\nat.C\ n 
  [C:\nat\rightarrow\SET]} \\

\recnat & P & Q & \0 &=& N \\

\recnat& P& Q& (\s\ x) &=& M\ x\ (\recnat\ N\ M\ x)
\end{array}
$

\medskip

In addition we add type connectives. $(+)$ stands for the type disjunction, 
and $(\times)$ for the pair type:

\medskip
$
\begin{array}{||l@{\ }c@{\ }r}
\multicolumn{3}{l}{
\hspace{-2mm}(+) :\SET\rightarrow\SET\rightarrow\SET}\\
\INL &:& A\rightarrow A+B~[A,B:\SET] \\
\INR &:& B\rightarrow A+B~[A,B:\SET] \\

\multicolumn{3}{l}{}\\[-3mm]
\multicolumn{3}{l}{
\hspace{-2mm}(\times) :\SET\rightarrow\SET\rightarrow\SET }\\
\PAIR &:& A\rightarrow B\rightarrow A\times B~[A,B:\SET]
\end{array}
$
\medskip

\noindent We write $(x,y)$ instead of $\PAIR~x~y$, and $(x_1,\dots,x_n)$ for
$(\dots (x_1,x_2),\dots,x_n)$.

We also need the empty type $\nzero$ (with no constructor):

\medskip
$
\nzero : \SET
$
\medskip

with which we can define $\EXIT$, its elimination rule, also known as
\emph{ex falsum quod libet} and the negation $\neg$:

\medskip

$
\begin{array}{||lcl}
\multicolumn{3}{l}{
\hspace{-2mm}\EXIT : \NNO\rightarrow A~[A:\SET]}\\[1mm]
\multicolumn{3}{l}{
\hspace{-2mm}\neg:\SET\rightarrow\SET}\\

\neg~A &=& A\rightarrow\nzero
\end{array}
$

\medskip

\noindent Notice that the constant $\EXIT$ has no computation rule.

The last type we need to define is $\none$, the unit type (\emph{i.e.} with only one trivial constructor), in other word the type ``true'':

\medskip
$
\begin{array}{||l@{\ }c@{\ }r}
\multicolumn{3}{l}{
\hspace{-2mm}\none : \SET }\\
\0 &:& \none
\end{array}
$
\medskip

\noindent Notice that $\0$ is polymorphic and is a constructor of both $\none$ and $\nat$.

 We can now start defining the more specific functions of our language. First comes
$(\leq)$. It decides if its first argument is less or equal to its second one.
Note that it returns either $\none$ or $\nzero$ which are types. This is an 
example of strong elimination, \emph{i.e} defining a predicate using a recursive
function.

\medskip

$
\begin{array}{||c@{~}c@{~}ccl}
\multicolumn{5}{l}{
\hspace{-2mm}(\leq) :\nat\rightarrow\nat\rightarrow \SET}\\

\0 &\leq & n &=& \none\\

(\s~x) & \leq & \0 &=& \nzero\\

(\s~x) & \leq & (\s~n) &=& x\leq n
\end{array}
$

\medskip

 Consequently we have the function $\PLEQ$ which proves essentially
that $(\leq)$ is a total ordering:

\medskip

$
\begin{array}{||l@{\ }c@{\ }ccl}
\multicolumn{5}{l}{
\hspace{-2mm}\PLEQ : \Pi x:\nat.\Pi n:\nat.(\s~x\leq n)+(n\leq x)}\\

\PLEQ & x & \0 &=& \INR~\0\\

\PLEQ &\0 &(\s~n) &=& \INL~\0\\

\PLEQ &(\s~x)&(\s~n) &=& \PLEQ~x~n
\end{array}
$

\medskip

 In order to write the proof of the shifting rule it is convenient to have a type
of vectors $\VEC~B~n$, which is intuitively $(\dots (\none\times B~\0)\dots)\times B~(n-1)$
and an access function of type

$\Pi n:\nat.\Pi x:\nat.(S~x\leq n)\rightarrow \VEC~B~n\rightarrow B~x$

Notice that this access function requires as an extra argument a proof that 
the index access is in the right range. To have such an access function is a nice
exercise in programming with dependent types.

This has to be seen as the type of finite approximations of proofs of 
$\Pi n:\nat. B~n$. And the access function is the respective elimination rule
(\emph{i.e.} a finite version of the forall elimination rule of natural deduction).

 The type of vectors $\VEC$ is defined recursively

\medskip

$
\begin{array}{||l@{\ }c@{\ }ccl}
\multicolumn{5}{l}{
\hspace{-2mm}\VEC: (\nat\rightarrow\SET)\rightarrow \nat\rightarrow \SET}\\
\VEC & B & \0 &=& \none \\

\VEC & B & (\s\ x) &=& (\VEC~B~x)\times B~x
\end{array}
$

\medskip

With $\VEC$ come two simple functions $\HEAD$ and $\TAIL$ accessing
respectively the two component of the pair (any non-$\0$-indexed vector
is a pair of an ``element'' and a shorter vector):

\medskip

$
\begin{array}{||l@{\ }c@{\ }c@{\ }cl}
\multicolumn{5}{l}{
\hspace{-2mm}\HEAD : \Pi x:\nat.(\VEC~B~(\s~x))\rightarrow B~x}\\

\HEAD & x & (v,u) &=& u \\

\multicolumn{5}{l}{} \\[-3mm]
\multicolumn{5}{l}{
\hspace{-2mm}\TAIL : \Pi x:\nat.(\VEC~B~(\s~x))\rightarrow \VEC~B~x}\\

\TAIL & x & (v,u) &=& v
\end{array}
$

\medskip

 In order to build the access function for type $\VEC$ (which is supposed
to extract the element of type $B~x$ from a vector of a length longer than $x$) we 
introduce a function $\getaux$ which shortens a vector of type $\VEC~B~n$ into
a vector of type $\VEC~B~x$ by removing the $n-x$ first elements. The reason why
such a function is useful is because we are trying to read the vector from the
inside to the outside. 

\medskip

$
\begin{array}{||l@{\ }ccl}
\multicolumn{4}{l}{
\hspace{-2mm}T:(\nat\rightarrow\SET)\rightarrow \SET}\\

T & P &=& \Pi k:\nat.P~(\s~k)\rightarrow P~k
\end{array}
$

\hfill

$
\begin{array}{||l@{\ }c@{\ }c@{\ }c@{\ }c@{\ }c@{\ }ccl}
\multicolumn{9}{l}{
\hspace{-2mm}\getaux : \Pi n:\nat.\Pi m:\nat.(n\leq m)\rightarrow \Pi P:\nat\rightarrow\SET.
T~P\rightarrow P~m\rightarrow P~n}\\

\getaux & 0 & 0 & p & P & h & v &=& v \\

\getaux & 0 & (\s~m) & p & P & h & v &=& \getaux~0~m~P~h~(h~m~v)\\

\getaux & (\s~n) & 0 & p & P & h & v &=& \EXIT~p \\

\getaux & (\s~n) & (\s~m) & p & P & h & v &=& \getaux~n~m~p~(\lambda x.P~(\s~x))~(\lambda x.h~(\s~x))~v
\end{array}
$

\medskip

As a consequence of the function $\getaux$ we can define in a rather simple way
the access function $\get$:

\medskip

$
\begin{array}{||l@{\ }c@{\ }c@{\ }c@{\ }c@{\ }ccl}
\multicolumn{8}{l}{
\hspace{-2mm}\get  : \Pi B:\nat\rightarrow\SET. \Pi n:\nat.\Pi x:\nat.(\s~x\leq n)\rightarrow \VEC~B~n\rightarrow B~x}\\

\get & B & n & x & p & v &=& \HEAD~x~(\getaux~(\s~x)~n~p~(\VEC~B)~\TAIL~v)
\end{array}
$

\medskip

We need the following result on the domain interpretation of 
this function $\get$. 
To simplify the notations we write $h$ instead of 
$\sem{h}{}$ if $h$ is a constant of the language. We also write 
$\NUM{l}$ for $\s^l~0$.

\begin{lemma}\label{extend}
Let $v\neq\bot, y\neq\bot$ and $B$ such that for any $l$, 
$B~(S^l~\TOP)\neq\bot$ and $B~\NUM{l}\neq\bot$ (in particular, 
$B\neq\bot$). If $x=\NUM{q}$ with $q<p$ then
$\get~\NUM{p}~x~0~v = \get~\NUM{p+1}~x~0~(v,y)$.
If $x = \s^q~\TOP$ with $q<p$ then $\get~\NUM{p}~x~0~v = \TOP$.
\end{lemma}
\begin{proof}
Let us prove that if $x=\NUM{q}$ with $q<p$ then
$\get~\NUM{p}~x~0~v = \get~\NUM{p+1}~x~0~(v,y)$. The proof of
the second part of the Lemma is similar.
It is proved by the following sequence of propositions
\begin{itemize}
\item If 
      $h=\sem{\lambda x.f~(\s~x)}{(f=h)}\neq\bot$ and $h~m~u=h~\TOP~u$ for
      any $m,u$,
      $q\leq p$, $t\neq\bot$, $v\neq\bot$ and 
      $P~(\s^l~\TOP)\neq\bot$ for any $l$ (in particular, $P\neq\bot$), 
      then $\getaux~\NUM{q}~\NUM{p}~t~P~v=
            (h~\TOP)^{p-q}~v$.

      This is proved by simple induction on $q$ and $p$. Using the definition
      of $\getaux$ together with Theorem \ref{model} and the fact that
      $P~(\s^l~\TOP)\neq\bot$ implies that 
      $\sem{\lambda f.f~(\s~x)}{(f=P)}~(\s^l~\TOP) = P~(\s^{l+1}~\TOP)
       \neq\bot$ for any $l$.
\item $\TAIL=\sem{\lambda x.f~(\s~x)}{(f=\TAIL)}\neq\bot$ and 
      $\TAIL~m~u=\TAIL~\TOP~u$.
      By Theorem \ref{model}.
\item If $B~(S^l~\TOP)\neq\bot$ and $B~\NUM{l}\neq\bot$, then
      for all $l$ $\VEC~B~(\s^l~\TOP)\neq\bot$.
      It is direct by induction on $l$ using the definition of 
      $\VEC$ and Theorem \ref{model}.
\item Finally
      $$
       \begin{array}{ll}
        \get~\NUM{p+1}~x~0~(v,y) & 
           = \HEAD~x~(\getaux~(\s~x)~\NUM{p+1}~0~(\VEC~B)~\TAIL~(v,y))\\
          &= \HEAD~x~((\TAIL~\TOP)^{p-q}~(v,y))\\
          &= \HEAD~x~((\TAIL~\TOP)^{p-q-1}~v)\\
          &= \HEAD~x~(\getaux~(\s~x)~\NUM{p}~0~(\VEC~B)~\TAIL~v)\\
          &= \get~\NUM{p}~x~0~v
       \end{array}
      $$
\end{itemize}
\end{proof}

 We can now introduce two functions $\brec$ and $\brecp$, defined in a mutual
recursive way. They define a slight generalisation of the double negation shift:

\medskip

$\begin{array}{||l@{\ }c@{\ }c@{\ }c@{\ }c@{\ }c@{\ }c@{\ }ccl}
\multicolumn{10}{l}{
\hspace{-2mm}\brec : \Pi B:\nat\rightarrow\SET. (\Pi n:\nat.\neg\neg B~n)\rightarrow \neg (\Pi n:\nat.B~n)\rightarrow \Pi n:\nat.\neg \VEC~B~n}\\

\multicolumn{10}{l}{
\hspace{-2mm}\brecp :  \Pi B:\nat\rightarrow\SET. (\Pi n:\nat.\neg\neg B~n)\rightarrow \neg (\Pi n:\nat.B~n)\rightarrow }\\

\multicolumn{10}{r}{
\Pi n:\nat.\VEC~B~n\rightarrow 
           \Pi x:\nat. (\s~x\leq n)+(n\leq x)\rightarrow B~x}\\

\brec & B & H & K & n & v &&&=&  K\ (\abs{x}{ \brecp\ B\ H\ K\ n\ v\ x\ (\PLEQ\ x\ n) })\\

\brecp & B & H & K & n & v & x & (\INL~p) &=& \get~B~n~x~p~v \\

\brecp & B & H & K & n & v & x & (\INR~p) &=& \EXIT~(H~n~(\lambda y.\brec~B~H~K~(\s~n)~(v,y)))
\end{array}
$

\medskip

 The program that proves Spector's double negation shift
$$
\Pi B:\nat\rightarrow\SET.(\Pi n:\nat.\neg\neg B~n)\rightarrow \neg \neg (\Pi n:\nat.B~n)
$$
is then $\lambda B. \lambda H.\lambda K.\brec~B~H~K~\0~\0$.

\section{Model of type theory and strong normalisation}

\subsection{Model}

We let $\POW$ be the collection of all subsets of $\DD$.
If $X\in\POW$ and $F:X\rightarrow\POW$ we define $\Pi(X,F)\in \POW$ by
$v\in\Pi(X,F)$ if and only if $u\in X$ implies $v~u\in {F(u)}$.

 A {\em totality predicate} on $\DD$ is a subset $X$ such that $\perp\notin X$ and $\TOP\in X$. 
We let $\TOT$ be the collection of all totality predicates.

\begin{lemma}
If $X\in\TOT$ and $F:X\rightarrow\TOT$ then $\Pi(X,F)\in\TOT$.
\end{lemma}

\begin{proof}
We have $\TOP\in X$. If $v\in\Pi(X,F)$  then  $v~\TOP\in F(\TOP)$ and so $v~\TOP\neq\perp$ and
$v\neq\perp$ hold. If $u\in X$ then $u\neq\perp$ so that $\TOP~u =\TOP\in F(u)$. This shows $\TOP\in \Pi(X,F)$.
\end{proof}

\begin{definition}\label{modeldef}
A {\em model} of type theory
is a pair $T,El$ with $T\in\TOT$  and $El:T\rightarrow\TOT$ satisfying the property:
if $A\in T$ and  $u\in El(A)$ implies $F~u\in T$ 
then $\FUN~A~F\in T$. Furthermore $El(\FUN~A~F) = \Pi(El(A),\lambda u.El(F~u))$.

If we have a collection of constants with typing rules $\vdash h:A$ we require also
$\SEM{A}\in T$ and $\SEM{h}\in El(\SEM{A})$. 

Finally, for a model of type theory
with universe $\SET$ we require also:
$\SET\in T$, $El(\SET)\subseteq T$ and $\FUN~A~F\in El(\SET)$ if $A\in El(\SET)$ and
$F~u\in El(\SET)$ for $u\in El(A)$.
\end{definition}

 The intuition is the following: $T\subseteq \DD$ is the collection of elements representing
types and if $A\in T$ the set $El~A$ is the set of elements of type $A$. The first
condition expresses that $T$ is closed under the dependent product operation. The last
condition expresses that $\SET$ is a type and that $El~(\SET)$ is a subset of $T$ which is
also closed under the dependent product operation.

 The next result states the soundness of the semantics w.r.t. the type system.

\begin{theorem}\label{sound}
  Let $\Delta$ be a context.
Assume that $\SEM{A}_{\rho}\in T$ and $\rho(x)\in El(\SEM{A}_{\rho})$ for $x{:}A$ in $\Delta$.
If $\Delta\vdash A$ then $\SEM{A}_{\rho}\in T$. If
$\Delta\vdash M{:}A$ then $\SEM{A}_{\rho}\in T$ and
$\SEM{M}_{\rho}\in El(\SEM{A}_{\rho})$.
\end{theorem}

\begin{proof}
Direct by induction on derivations, using Theorem \ref{model}
and Corollary \ref{subst}. For instance, we justify the application rule. We have
by induction $\sem{N}{\rho}\in \Elem(\FUN~\sem{A}{\rho}~\sem{\lambda x.B}{\rho})$ and
$\sem{M}{\rho}\in El(\sem{A}{\rho})$. It follows that we have
$$
\sem{N~M}{\rho} = \sem{N}{\rho}~\sem{M}{\rho} \in El(\sem{\lambda x.B}{\rho}~\sem{M}{\rho})
$$
Since $El(\sem{A}{\rho})\in\TOT$ we have $\sem{M}{\rho}\neq\perp$. Hence by
Theorem \ref{model} and Corollary \ref{subst} we have
$$
\sem{\lambda x.B}{\rho}~\sem{M}{\rho} = \sem{B}{\rho,x = \sem{M}{\rho}} = \sem{B[M]}{\rho}
$$
and so $\sem{N~M}{\rho}\in El(\sem{B[M]}{\rho})$ as expected.
\end{proof}

\subsection{Construction of a model}

\begin{theorem}\label{total}
The filter model $\DD$ of UPL can be extended to a model $T\in\TOT,~El:T\rightarrow\TOT$.
\end{theorem}

\begin{proof}
 The 
main idea is to define the pair $T,El$ in two inductive steps, using Lemma \ref{partition}
to ensure the consistency of this definition. We define first $T_0,El$. We have $\TOP\in T_0$ and
$\TOP\in El(A)$ if $A\in T_0$. Furthermore, we have

\begin{itemize}

\item $\nzero\in T_0$

\item $\none\in T_0$ and $\ZERO\in El(\none)$

\item $\nat\in T_0$ and $\ZERO \in El(\nat)$ and $\SUCC~x\in El(\nat)$ if
$x\in El(\nat)$

\item $A+B\in T_0$ if $A,B\in T_0$ and $\INL~x\in El(A+B)$ if $x\in El(A)$
and $\INR~y\in El(A+B)$ if $y\in El(B)$

\item $A\times B\in T_0$ if $A,B\in T_0$ and $(x,y)\in El(A\times B)$ if $x\in El(A)$
and $y\in El(B)$

\item $\FUN~A~F\in T_0$ if $A\in T_0$ and $F~x\in T_0$ for $x\in El(A)$. Furthermore
$w\in El(\FUN~A~F)$ if $w~x\in El(F~x)$ whenever $x\in El(A)$

\end{itemize}

 We can then define $T\supseteq T_0$ and the extension $El:T\rightarrow \TOT$ by the same
conditions extended by one clause

\begin{itemize}

\item $\nzero\in T$

\item $\none\in T$ and $\ZERO\in El(\none)$

\item $\nat\in T$ and $\ZERO \in El(\nat)$ and $\SUCC~x\in El(\nat)$ if
$x\in El(\nat)$

\item $A+B\in T$ if $A,B\in T$ and $\INL~x\in El(A+B)$ if $x\in El(A)$
and $\INR~y\in El(A+B)$ if $y\in El(B)$

\item $A\times B\in T$ if $A,B\in T$ and $(x,y)\in El(A\times B)$ if $x\in El(A)$
and $y\in El(B)$

\item $\FUN~A~F\in T$ if $A\in T$ and $F~x\in T$ for $x\in El(A)$. Furthermore
$w\in El(\FUN~A~F)$ if $w~x\in El(F~x)$ whenever $x\in El(A)$

\item $\SET\in T$ and $El(\SET) = T_0$

\end{itemize}

 The definition of the pair $T,El$ is a typical example of an {\em inductive-recursive} definition:
we define simulatenously the subset $T$ {\em and} the function $El$ on this subset.
The justification of such a definition is subtle, but it is 
standard \cite{Aczel,Beeson,Scott1}. 
It can be checked by induction that $T\in \TOT$ and $El(A)\in\TOT$ if $A\in T$.
The next subsection proves that $\sem{h}{}\in El~(\sem{A}{})$ if $\vdash h{:}A$ is
a typing rule for a constant $h$. 
\end{proof}

\subsection{Strong normalisation via totality}

 It is rather straightforward to check that we have $\sem{h}{}\in El(\sem{A}{})$ for all
the constants $h:A$ that we have introduced except the last two constants $\brec$ and $\brecp$. 
For instance $\sem{\EXIT}{}\in El(\nzero\rightarrow A)$ for
any $A\in T$ since $El(\nzero) = \{\TOP\}$ and $\sem{\EXIT}{}~\TOP = \TOP$ is in
$El(A)$. To check $\sem{h}{}\in El(\sem{A}{})$ is more complex for the last two functions. 

\begin{theorem}
For all constants $h:A$ that we have introduced, we have $\sem{h}{}\in El(\sem{A}{})$.
\end{theorem}

\begin{proof}
To simplify the notations we write $h$ instead of $\sem{h}{}$ if $h$ is a constant of
the language, and we say simply that $h$ is total instead of $h\in El(A)$.
The only difficult cases are for the constants ${\brec}$ and ${\brecp}$.
It is the only place where we use classical reasoning. We only write the 
proof for ${\brec}$, the case of ${\brecp}$ is similar.

 Assume that ${\brec}$ is not total. We can then find total elements $B\in El(\nat\rightarrow\SET)$, 
$H\in El(\FUN~\nat~(\lambda x.\neg\neg~(B~x)))$, 
$K\in El(\neg~(\FUN~\nat~B))$, $n\in El(\nat)$ and 
$v\in El(B~n)$ such that $\brec~B~H~K~n~v$ does not belong to $El(\nzero) = \{\top\}$.
Since 
$$\brec~B~H~K~n~v = K~(\lambda x.\brecp\ B\ H\ K\ n\ v\ x\ (\PLEQ\ x\ n))$$
and $K$ is total, there exists $x\in El(\nat)$ such that
$\brecp\ B\ H\ K\ n\ v\ x\ (\PLEQ\ x\ n)$ is not total at type $B~x$. Given the definition of
$\brecp$ this implies that $\PLEQ\ x\ n$ is of the form $\INR~h$. It follows from the definition
of $\PLEQ$ that $n$ is of the form $\NUM{p}$. Furthermore 
$$\brecp\ B\ H\ K\ n\ v\ x\ (\PLEQ\ x\ n) = 
  \EXIT~(H~\NUM{p}~(\lambda y.\brec~H~K~\NUM{p+1}~(v,y)))$$
is not total. Since $H$ is total, there exists 
$y_p\in El~(B~\NUM{p})$ such that
$\brec~B~H~K~\NUM{p+1}~(v,y_p)$ is not total. Reasoning
in the same way, we see that there exists $y_{p+1}\in El~(B~\NUM{p+1})$ such that
$\brec~B~H~K~\NUM{p+2}~(v,y_p,y_{p+1})$ is not total. Thus we build a sequence
of elements $y_m\in El~(B~\NUM{m})$ for $m\geq p$ such that, for any $m$ 
$$\brec~B~H~K~\NUM{m}~(v,y_p,\dots,y_{m-1})\neq \TOP$$

 Consider now an element $x= \NUM{q}$. For $m>q$ we have $\s~x\leq \NUM{m} = \none$
and we take $f~x$ to be $\get~\NUM{m}~x~0~(v,y_p,\dots,y_{m-1})$. This is well defined 
since we have for $m_1,m_2>q$ by Lemma \ref{extend}
$$\get~B~\NUM{m_1}~x~0~(v,y_p,\dots,y_{{m_1}-1}) = 
  \get~B~\NUM{m_2}~x~0~(v,y_p,\dots,y_{{m_2}-1})$$
We take also $f~(\s^q~\top) = \TOP$. This defines
a total element $f$ in $El~(\FUN~\nat~(\lambda x.El~(B~x)))$. Since
$K$ is total, $K~f$ is total and belongs to $El~(\nzero) = \{\top\}$. Hence $K~f = \TOP$. Since
$\TOP$ is a finite element of $\DD$ we have by continuity $K~f_0 = \TOP$ for some
finite approximation $f_0$ of $f$. In particular there exists $m$ such that if
$g_m~(\s^q~0) = f~(\s^q~0)$  and $g_m~(\s^q~\TOP) = f~(\s^q~\TOP)$,
for all $q<m$, then $K~g_m = \TOP$. If we define
$$
g_m~x = \brecp\ B\ H\ K~\NUM{m}~(v,y_p,\dots,y_{m-1})\ x\ (\PLEQ\ x\ \NUM{m})
$$
we do have $g_m~(\s^q~0) = f~(\s^q~0)$  and 
$g_m~(\s^q~\TOP) = f~(\s^q~\TOP)$ for all $q<m$. Hence $K~g_m = \TOP$. But then
$$\brec~B~H~K~\NUM{m}~(v,y_p,\dots,y_{m-1}) = K~g_m = \TOP$$
which contradicts the fact that the element
$\brec~B~H~K~\NUM{m}~(v,y_p,\dots,y_{m-1})$ is {\em not} total.
\end{proof}

Like in \cite{Berger}, it is crucial for this argument that we are using a domain model.
These constants make also the system proof-theoretically strong, at least the strength
of second-order arithmetic.

\begin{corollary}\label{sem}
If $\vdash A$ then $\SEM{A}\neq\perp$. If $\vdash M:A$ then $\SEM{M}\neq\perp$.
\end{corollary}

\begin{proof}
If $\vdash A$  we have by Theorem \ref{sound} that $\SEM{A}\in T$. By Theorem \ref{total}
we have $T\in\TOT$. Hence $\SEM{A}\neq\perp$.
Similarly, if $\vdash M:A$ we have by Theorem \ref{sound} that $\SEM{A}\in T$ and
$\SEM{M}\in El(\SEM{A})$. By Theorem \ref{total} we have $T\in\TOT$ and $El(\SEM{A})\in\TOT$.
Hence $\SEM{A}\neq\perp$ and $\SEM{M}\neq\perp$.
\end{proof}

 By combining Corollary \ref{sem} with Theorem \ref{SN} we get

\begin{theorem}
If $\vdash A$ then $A$ is strongly normalisable. If $\vdash M:A$ then $M$
is strongly normalisable.
\end{theorem}

\section*{Conclusion}
 We have built a filter model $\DD$ for an untyped calculus having the property that a term
is strongly normalisable whenever its semantics is $\neq\perp$, and then used this to give
various {\em modular} proofs of strong normalization. While each part uses
essentially variation on standard materials, our use of filter models 
seems to be new and can be seen as an application of computing science
to proof theory. It is interesting that we are naturally lead in this way to consider a domain with a top element.
 We have shown on some
examples that this can be used to prove strong normalisation theorem in a modular way,
essentially by reducing this problem to show the soundness of a semantics over the 
domain $\DD$. 
There should be no problem to use our model to give a simple normalisation proof of
system F extended with bar recursion. It is indeed direct that totality predicates are closed
under arbitrary non empty intersections. By working in the $\DD$-set model 
over $\DD$ \cite{Streicher, Altenkirch}, one should be able to get also
strong normalisation theorems for various impredicative type theories extended with bar recursion.

 For proving normalisation for {\em predicative} type systems, the use of the model $\DD$
is proof-theoretically too strong: the totality predicates are sets of filters, that are 
themselves sets of formal neighbourhoods, and so are essentially third-order objects.
For applications not involving strong schemas
like bar recursion, it is possible however to work instead only with the definable elements of the
set $\DD$, and the totality predicates become second-order objects, as usual. It 
is then natural to extend our programming language with an extra element $\top$ that 
plays the role of a top-level error.
As suggested also to us by Andreas Abel,
it seems likely that Theorem \ref{SNinter} has a purely combinatorial proof, similar
in complexity to the one for simply typed $\lambda$-calculus. He gave such 
a proof for a reasonable subsystem in \cite{Abel}. 

 A natural extension of this work would be also to state and prove a {\em density} theorem
for our denotational semantics, following \cite{density}. The first step would be to define
when a formal neighbourhood is of a given type.

 In \cite{Bakel,Pottinger}, for untyped $\lambda$-calculus without constants, it is proved that
a term $M$ is strongly normalizing if {\em and only if} $\SEM{M}\neq\perp$. This does not hold
here since we have for instance $\0\ \nat$ strongly normalizing, but $\SEM{\0\ \nat} = \perp$.
However, it may be possible to find a natural subset of terms $M$ for which the
equivalence between  $M$ is strongly normalizing and $\SEM{M}\neq\perp$ holds.
Additionally, Colin Riba showed this result for a system where the 
neighbourhoods
are closed by union but were the rewrite rules are weaker \cite{Riba}.

 Most of our results hold without the hypotheses that the rewrite rules are mutually
disjoint. We only have to change the typing rules for a constant $f$ in Figure \ref{f:typeinter} by
the uniform rule:
$\Gamma\vdash _{\mathcal{M}} f:U_1\rightarrow\ldots\rightarrow U_k\rightarrow V$ if 
{\em for all} rules $f\ p_1\ldots p_k = M$ and {\em for all} $W_1,\ldots,W_n$ such that
$p_i(W_1,\ldots,W_n) = U_i$ we have $\Gamma,x_1:W_1,\dots,x_n:W_n\vdash _{\mathcal{M}} M:V$.
(This holds for instance trivially in the special case where no rules for $f$ matches
$U_1,\ldots,U_n$.) For instance, we can add a constant $+$ with rewrite rules
$$
\begin{array}{ccccc}
+ & n & \0 & = & n \\
+ & \0 & n & = & n \\
+ & n & (\s\ m) & = & \s\ (+\ n\ m) \\
+ & (\s\ n) & m & = & \s\ (+\ n\ m) 
\end{array}
$$
and Theorem \ref{SN} is still valid for this extension. 

\section*{Acknowledgement}

 Thanks to Mariangiola Dezani-Ciancaglini for the reference to the paper \cite{Bakel}.
The first author wants also to thank Thomas Ehrhard for reminding him about proofs of
strong normalisation via intersection types.

\end{document}